\documentclass[journal]{IEEEtran}

\usepackage[pdftex]{graphicx}
\usepackage[T1]{fontenc}
\usepackage{lmodern}
\usepackage{textcomp}
\usepackage{latexsym}
\usepackage{cite}
\usepackage{bm}
\usepackage{amsmath}
\usepackage{amsfonts}
\usepackage[subrefformat=parens]{subcaption}
\usepackage{multirow}
\usepackage{booktabs}
\usepackage{array}
\usepackage{tabularx}
\usepackage{arydshln}

\begin{document}
\title{Accurate Radar-Based Detection of Sleep Apnea Using Overlapping Time-Interval Averaging}

\author{Kodai~Hasegawa, Shigeaki~Okumura, Hirofumi~Taki,~\IEEEmembership{Member, IEEE}, Hironobu~Sunadome, Satoshi~Hamada, Susumu~Sato, Kazuo~Chin, and Takuya~Sakamoto, ~\IEEEmembership{Senior Member, IEEE}
  \thanks{K.~Hasegawa and T.~Sakamoto are with the Graduate School of Engineering, Kyoto University, Kyoto 615-8510, Japan.}
  \thanks{S.~Okumura and H.~Taki are with MaRI Company Ltd., Kyoto 600-8815, Japan.}
  \thanks{H.~Sunadome, S.~Hamada, and S.~Sato are with the Graduate School of Medicine, Kyoto University, Kyoto 606-8501, Japan.}
  \thanks{K.~Chin is with the Graduate School of Medicine, Kyoto University, Kyoto 606-8501, Japan, and also with the School of Medicine, Nihon University, Itabashi-ku, Tokyo 173-8610, Japan.}}
\markboth{}%
{Hasegawa \emph{et al.}: Accurate Radar-Based Detection of Sleep Apnea Using Overlapping Time-Interval Averaging}

\maketitle
\begin{abstract}
  Radar-based respiratory measurement is a promising tool for the noncontact detection of sleep apnea.
  Our team has reported that apnea events can be accurately detected using the statistical characteristics of the amplitude of respiratory displacement.
  However, apnea and hypopnea events are often followed by irregular breathing, reducing the detection accuracy.
  This study proposes a new method to overcome this performance degradation by repeatedly applying the detection method to radar data sets corresponding to multiple overlapping time intervals.
  Averaging the detected classes over multiple time intervals gives an analog value between 0 and 1, which can be interpreted as the probability that there is an apnea event.
  We show that the proposed method can mitigate the effect of irregular breathing that occurs after apnea / hypopnea events, and its performance is confirmed by experimental data taken from seven patients.
\end{abstract}


\IEEEpeerreviewmaketitle

\section{Introduction}
\IEEEPARstart{I}{n} recent years, the number of patients with sleep apnea syndrome (SAS) has been increasing, with 936 million people worldwide suffering from SAS~\cite{sleepapnea}.
SAS is associated with daytime sleepiness, impaired cognitive function, and complications such as hypertension, cardiovascular disease, and stroke \cite{sleep,complicate,card}, making early detection crucial.
The diagnosis is typically based on polysomnography (PSG).
PSG is known as a gold standard for providing an accurate diagnosis using contact sensors, cameras, and microphones to detect apnea and hypopnea.
However, the use of such sensors for an extended period during sleep can be a burden on patients and can disrupt their sleep \cite{PSG}.

To reduce this burden, radar-based non-contact methods have been explored.
These techniques detect apnea by measuring body movements caused by respiration.
One common approach is threshold-based detection, which identifies apnea events when the displacement amplitude exceeds a threshold.
However, setting an appropriate threshold requires training data that include normal breathing patterns \cite{nonmachine1, nonmachine2, nonmachine3}.
Machine learning offers another approach \cite{machine1, machine2, machine3, machine4}, but achieving high detection accuracy requires a large volume of training data.
Both approaches face hurdles when applied outside of controlled laboratory settings.

To address the aforementioned challenges, Koda et al. \cite{EM} proposed an effective method using the expectation maximization (EM) algorithm, focusing on the statistical distribution of the displacement amplitude.
This method achieves high accuracy without requiring data for the training of machine learning models or predefined thresholds used in threshold-based methods.
Despite this, the detection accuracy of the method can deteriorate depending on the chosen time intervals for applying the EM algorithm, because the accuracy depends on whether these intervals include irregular respiration following apnea events.

In this study, we build on the method proposed in \cite{EM} by applying the EM algorithm to multiple overlapping time intervals and averaging the resulting classifications.
To evaluate the performance of the proposed method, we apply the conventional and proposed methods to radar data taken from experiments on seven SAS patients and demonstrate the effectiveness of the proposed method in improving detection accuracy.

\section{Apnea Detection Using Radar and the EM Algorithm}
In scenarios where the radar captures a single dominant echo from the human body, the body displacement can be estimated as $d'(t) = ({\lambda}/{4\pi})\mathrm{unwrap}\{\angle s(t)\},$ where $\lambda$ is the wavelength, $s(t)$ is the signal reflected from the target, $\mathrm{unwrap} \{ \cdot \}$ is the phase unwrapping operator, and $\angle$ represents the phase of a complex number.
To extract only the respiratory displacement $d(t)$ with a period between 3.0 and 5.0 s, a bandpass filter is applied to $d'(t)$ as follows: $d(t) = [d'(t)-d'*h_1(t)]*h_2(t)$, where $*$ denotes the convolution operator.
In this study, $h_1(t)$ and $h_2(t)$ represent a rectangular window with a window length of 6.0 s and a Hann window with a window length of 1.1 s, respectively.
The respiratory displacement amplitude $\bar{d}(t)$ is given by
\begin{align}\label{eq:amp}
  \bar{d}(t) = \sqrt{\frac{1}{T_\mathrm{a}}\int_{t-\frac{T_\mathrm{a}}{2}}^{t + \frac{T_\mathrm{a}}{2}} \left|d(t') \right|^2 \mathrm{d} t'},
\end{align}
where we set $T_\mathrm{a} = 5.0\,\mathrm{s}$.
In the method described in \cite{EM}, the amplitude of the respiratory displacement $\bar{d}(t)$ within a specific time interval $[t', t'']$ is considered a random variable.
Let us assume that during normal breathing, $\bar{d}(t)$ obeys a Gaussian distribution with a mean $\mu_1$ and a variance $\varSigma_1$, while during apnea events, $\bar{d}(t)$ follows another Gaussian distribution with a mean $\mu_2$ and a variance $\varSigma_2$, where $\mu_1 > \mu_2$.
We assume that the measured data consist of either (a) normal breathing only or (b) a combination of normal breathing and apnea events.
The statistical distribution $G(\bar{d})$ of $\bar{d}(t)$ for $t'\leq t \leq t''$ can be modeled as a Gaussian mixture model $G(\bar{d}|\bm{\pi},\bm{\mu},\bm{\varSigma})$ expressed as
\begin{align}
  G(\bar{d}|\bm{\pi},\bm{\mu},\bm{\varSigma}) = \sum_{k = 1}^{K}\pi_{k} \mathcal{N}(\bar{d}|\mu_{k}, \varSigma_{k}),
\end{align}
where the number of Gaussian components is fixed to be $K=2$, and $\mathcal{N}(\cdot|\mu,\varSigma)$ denotes the normal distribution with a mean $\mu$ and a variance $\varSigma$.
We define the parameter vectors $\bm{\mu}=[\mu_1, \mu_2]^{\mathrm{T}}$, $\bm{\varSigma}=[\varSigma_1, \varSigma_2]^{\mathrm{T}}$, and $\bm{\pi}=[\pi_1, \pi_2]^{\mathrm{T}}$, where $\bm{\pi}$ represents the mixing ratios and satisfies $\pi_1 + \pi_2 = 1$.

Using the EM algorithm, we obtain the optimal parameters $\bm{\pi}^{*}$, $\bm{\mu}^{*}$, and $\bm{\varSigma}^{*}$ with hidden variables $\gamma_{1}$ and $\gamma_2$.
Note that the EM algorithm is applied to the histogram of $\bar{d}(t)$ over the time interval $[t', t'']$.
Given that the median duration of apnea is 25.0~s for severe patients \cite{length}, the time interval $T = t'' - t'$ is set at 60.0~s throughout this study.
As a result, the time interval $T$ is expected to contain both normal breathing and an apnea event; i.e., we can exclude data that include only apnea.
Finally, we obtain a temporary label $L(t)$ as $L(t) = 1$ if an apnea event is detected at time $t$ and $L(t) = 0$ otherwise.
By applying the EM algorithm to the histogram of $\bar{d}(t)$, we estimate the label $L(t)$ as
\begin{align}\label{eq:label_multi}
  L(t) =
          \begin{cases}
              1 & (\gamma_{1}(t) \leq \gamma_{2}(t),\,\,\,\mathrm{and}\,\,\, {\mu_{2}^{*}}/{\mu_{1}^{*}}
              \leq \beta) \\
              0        & (\mathrm{otherwise}),
          \end{cases}
\end{align}
where $\beta=0.7$ is a ratio threshold.
Condition $\gamma_{1}(t) \leq \gamma_{2}(t)$ is used to detect the time $t$ at which an apnea event occurs.
Although the study \cite{EM} reports that this method is able to detect apnea events accurately, the detection accuracy depends on the selection of the time interval (i.e., the selection of $t'$ and $t''$) as explained in the next section.

\section{Apnea Detection Using Overlapping Time-Interval Averaging}
\begin{figure}[htbp]
  \centering

  \begin{subfigure}{0.45\textwidth}
    \centering
    \includegraphics[width=\linewidth]{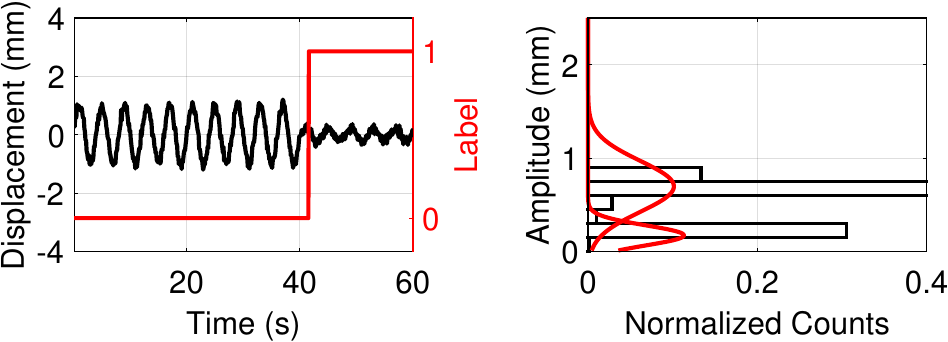}
    \caption{}
  \end{subfigure}

  \begin{subfigure}{0.45\textwidth}
    \centering
    \includegraphics[width=\linewidth]{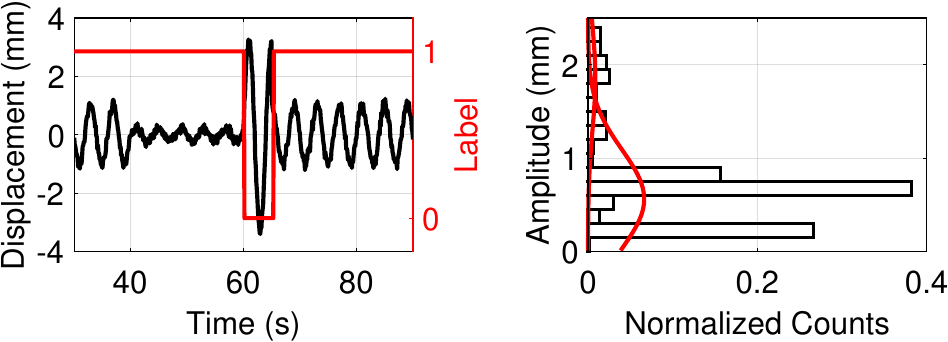}
    \caption{}
  \end{subfigure}
  \begin{subfigure}{0.45\textwidth}
    \centering
    \includegraphics[width=\linewidth]{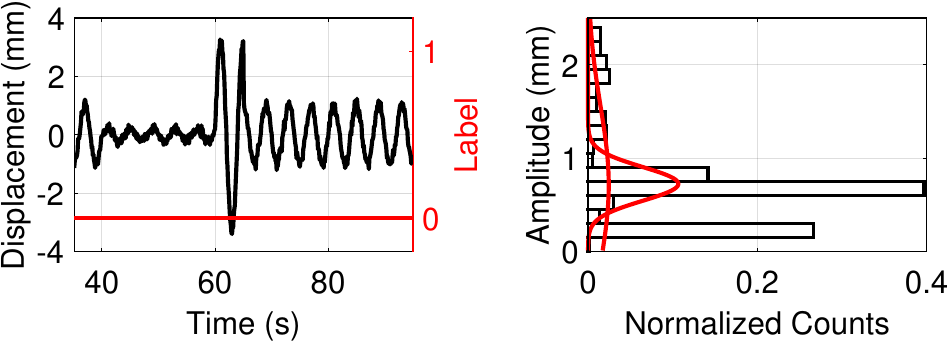}
    \caption{}
  \end{subfigure}
  \begin{subfigure}{0.45\textwidth}
    \centering
    \includegraphics[width=\linewidth]{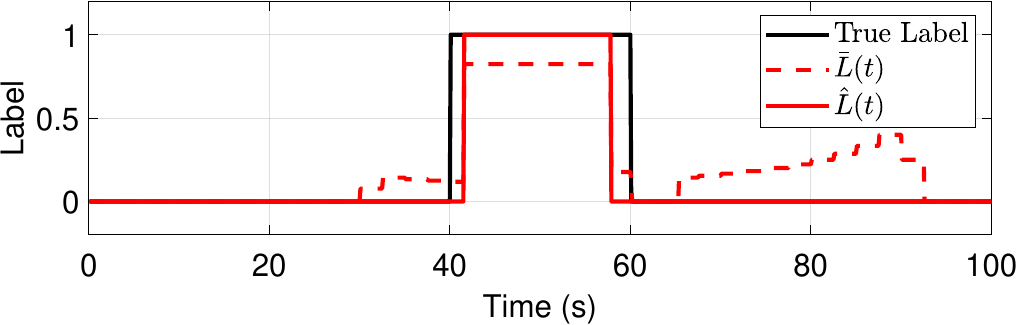}
    \caption{}

  \end{subfigure}

  \caption{Simulated sinusoidal displacement $d(t)$ for $t' \leq  t \leq t''$,  temporary labels $L(t)$, amplitude histogram $G(\bar d)$, and the estimated Gaussian mixture model
$G(\bar{d}|\bm{\pi}^{*},\bm{\mu}^{*},\bm{\varSigma}^{*})$ obtained using the EM algorithm within three different time intervals (a)--(c). Panel (d) shows the apnea detection results using overlapping time-interval averaging.}
  \label{fig:window}
\end{figure}

We often observe irregular body movements after each apnea event.
The detection accuracy of the method \cite{EM} decreases when such movements occur within the time interval $[t', t'']$ used to generate a histogram.
Consider a simulated sinusoidal displacement $d(t)$ with a period of 4.0 s, which represents (1) normal breathing for 40 s, (2) apnea/hypopnea for 20 s, (3) irregular movement for $T_\mathrm{m}=5$ s, and (4) normal breathing again for 35 s, where the amplitude values are set to 1.0, 0.1, and $d_\mathrm{m}=3.3$ mm for normal breathing, an apnea event, and irregular movement, respectively.
The left panels of Fig.~\ref{fig:window} (a)--(c) show $d(t)$ (black) and estimated temporary labels $L(t)$ (red) whereas the right panels show the amplitude histogram $G(\bar{d})$ of displacement (black) and the estimated Gaussian mixture model $G(\bar{d}|\bm{\pi}^{*},\bm{\mu}^{*},\bm{\varSigma}^{*})$ (red).
The method from \cite{EM} was applied to the displacement $d(t)$ over the intervals $[t', t'']=[0, 60\;\mathrm{s}]$, $[t', t'']=[30\; \mathrm{s}, 90\; \mathrm{s}]$, and $[t', t'']=[35\;\mathrm{s}, 95\;\mathrm{s}]$ in panels (a), (b), and (c), respectively.

In Fig.~\ref{fig:window} (a), the method from \cite{EM} correctly detects apnea (40--60 s).
In contrast, in Fig.~\ref{fig:window} (b), irregular movement negatively affects the performance of the method, resulting in the misclassification of normal breathing as an apnea event.
Furthermore, in Fig.~\ref{fig:window} (c), owing to irregular movement, the apnea event was not detected at all.
Therefore, detection performance is believed to strongly depend on the selection of the time interval $[t',t'']$; even if the data contain an apnea event, it may not be correctly detected without selecting an appropriate time interval.

In the proposed method, to determine whether an apnea event occurs at time $t$, we apply the method \cite{EM} to multiple overlapping intervals $[t',t'']$ containing $t$, which means $t'\leq t\leq t''$, regardless of whether the time interval includes irregular movements.
Assuming the total time length is $T_0$, data sets are generated for time intervals $I_1=[t_1, t_1+T], I_2=[t_2, t_2+T], \cdots, I_N = [t_N, t_N+T]=[T_0-T,T_0]$, where $t_n=(n-1)\Delta t$ $(n=1,2,\cdots,N)$, $\Delta t=2.5$ s is the sliding time step and $N$ is the number of time intervals.
The method from \cite{EM} is applied to all time intervals $I_n$ that satisfy $t_n\leq t\leq t_{n} + T$, and the temporary labels $L_n(t)$ are obtained.

\begin{figure}[tb]
  \centering
  \includegraphics[width =0.9\linewidth]{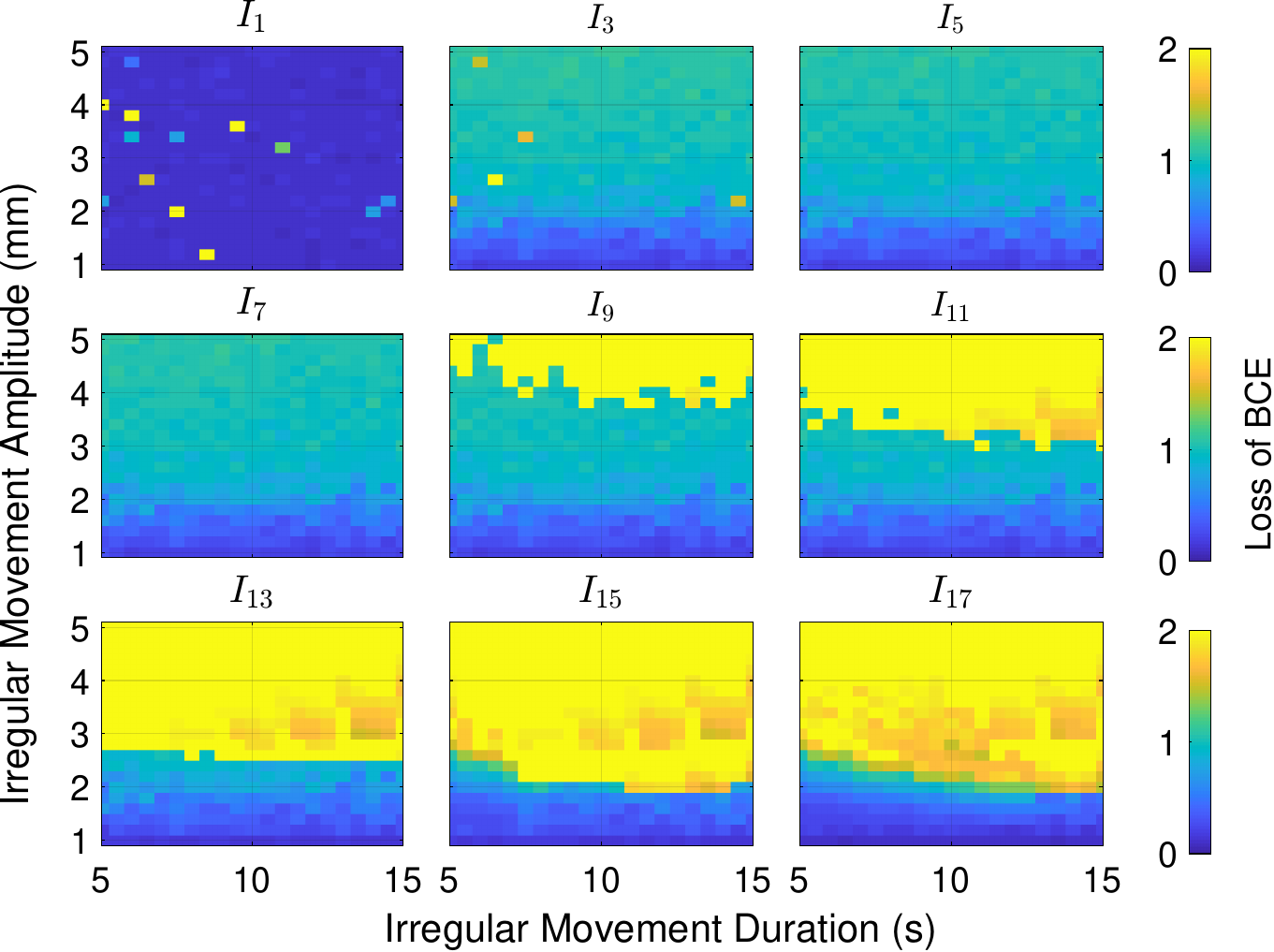}
  \caption{Loss of BCE for various irregular movement parameters.}
  \label{fig:bce}
\end{figure}

Next, we evaluate the effect of irregular movements with an amplitude $d_\mathrm{m}$ and a duration $T_\mathrm{m}$.
Fig.~\ref{fig:bce} shows the accuracy in classifying radar data for the time interval $I_n$ into normal breathing and apnea events using the EM algorithm, for various amplitudes $d_\mathrm{m}$ and durations $T_\mathrm{m}$ of irregular movements.
Here, accuracy refers to the degree of agreement between the hidden variables $\gamma_{n,2}(t)$ estimated using the EM algorithm and the true labels within $I_n$ in terms of the binary cross-entropy (BCE) loss, where a small loss of BCE indicates high accuracy.
We assume the same simplified model as used in Fig.~\ref{fig:window} for parameter ranges 1.0 mm $\leq d_\mathrm{m}\leq$ 5.0 mm and 5 s $\leq T_\mathrm{m}\leq$ 15 s.
These ranges are based on the average duration of arousals of 12.6 s (SD = 1.7) after apnea and 9.9 s (SD = 2.4) after hypopnea \cite{arousal}.
Fig.~\ref{fig:bce} shows the loss of BCE, where we see that as $d_\mathrm{m}$ increases, the loss of BCE increases for a large $T_\mathrm{m}$, deteriorating the accuracy.
However, the loss of BCE becomes small for most of the intervals within the parameter range of 1.0 mm $\leq d_\mathrm{m}\leq$ 3.5 mm and 5 s $\leq T_\mathrm{m}\leq$ 15 s, suggesting that averaging the classification labels over multiple time intervals is the key to improving the accuracy of the classification.

On the above basis, the temporary labels $L_n(t)$ are averaged to produce $\bar{L}(t)$ as
\begin{align}\label{eq:lp}
  \bar{L}(t) = \frac{1}{N'(t)}\sum_{n: t\in I_n}L_n(t),
\end{align}
where the summation is performed over all $n$ satisfying $t\in I_n$, and $N'(t)$ denotes the number of indices $n$ satisfying $t\in I_n$.
The resultant label $\bar{L}(t)$ takes a value between 0 and 1, which we regard as the probability of the existence of an apnea event at time $t$.
Next, $\bar{L}(t)$ is binarized using a threshold $L_\mathrm{th}$ as
\begin{align}\label{eq:lap}
  \hat{L}(t) =
          \begin{cases}
              1 & (\bar{L}(t) \geq L_\mathrm{th}) \\
              0        & (\mathrm{otherwise}).
          \end{cases}
\end{align}
Finally, we remove candidates for apnea even when $\hat{L}(t)=1$ if the duration of the apnea is less than 10 s, which means we substitute $\hat{L}(t)\leftarrow 0$.
This step is based on an American Academy of Sleep Medicine guideline, which defines apnea as a complete cessation of airflow for more than 10 s \cite{PSG}.
Fig.~\ref{fig:window} (d) shows the apnea detection results $\hat{L}(t)$ obtained using the proposed method with a threshold $L_\mathrm{th}=0.60$, revealing a clear improvement in the detection of sleep apnea events relative to the original method \cite{EM}.

\section{Experimental Performance Evaluation}
\begin{figure}[tb]
  \centering
  \includegraphics[width=0.8\columnwidth]{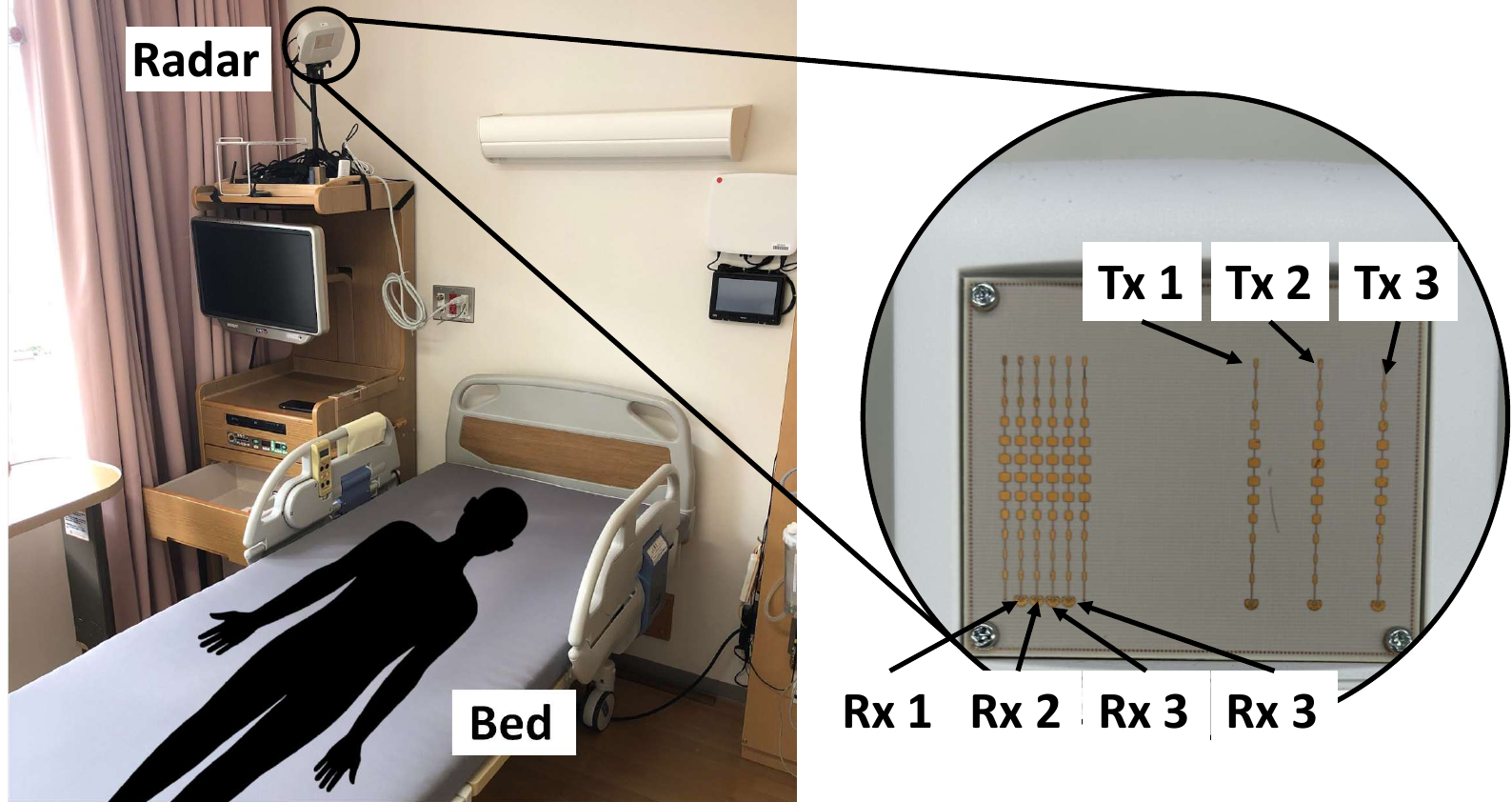}
  \caption{Experimental setup and millimeter-wave array radar.}
  \label{fig:flow}
\end{figure}

To evaluate the performance of the proposed method, experiments were conducted with seven patients having SAS symptoms on a hospital bed, as shown in Fig.~\ref{fig:flow}.
As detailed in Table~\ref{tab:1}, we used a millimeter-wave frequency-modulated continuous wave (FMCW) radar sysytem having a central frequency of 79 GHz and a bandwidth of 3.9 GHz.
The antenna array comprised three transmitters (Tx) and four receivers (Rx).
Each patient was monitored through polysomnography (PSG) during our radar measurement, and PSG recordings were used as reference for the performance evaluation of our radar-based methods.
The performance of the proposed method was evaluated using (1) the apnea hypopnea index (AHI), which is a diagnostic criterion defined as the number of apnea events per hour, and (2) the score $\mathrm{F}_1$, which represents the accuracy of the temporal alignment of the label.

\begin{table}[tb]
  \centering
  \caption{Radar parameters and patient details}
  \label{tab:1}
  \begin{tabular}{c|c||c|c}
    \toprule
    Parameter               & Value   & Patient No.& AHI  \\ \midrule
    Modulation              & FMCW    &1 &25.6\\
    Central frequency        & 79.0 GHz&2 &41.2\\
    Bandwidth               & 3.9 GHz &3 &15.2 \\
    No. of Tx elements      & 3       &4 &15.1\\
    No. of Rx elements      & 4       &5 &5.2\\
    Range resolution        & 43.0 mm &6 &13.6\\
    Sampling frequency      & 10 Hz   &7 &3.7\\
    \bottomrule
  \end{tabular}
\end{table}

\subsubsection{Use of the same threshold for all patients}
First, the performance of the detection of apnea was evaluated with the same threshold applied to all seven patients, which is henceforth referrd to as proposed method A.
Empirically, we found that the average F$_1$ score was maximized when we set $L_\mathrm{th}=0.60$.
Table~\ref{tab:ahi} gives the AHI error and the F$_1$ score when using the existing method \cite{EM} and the proposed method with $L_\mathrm{th} = 0.60$.
We see that with the proposed method, the average AHI error across all patients is reduced by a factor of 1.6, while the average F$_1$ score improves by a factor of 1.2.
In particular, for patients 1 and 2, the F$_1$ score exceeds 0.70, suggesting highly accurate detection.

\begin{table}[tb]
  \centering
  \caption{AHI error and $\mathrm{F}_1$ score estimated using conventional and proposed methods (Prop. A: same threshold, Prop. B: patient-specific threshold).}
  \small
  \renewcommand{\arraystretch}{1.2}
  \begin{tabular}{|c||c|c|c|} \hline
    Patient & Conv. & Prop. A & Prop. B \\\hline
    \multicolumn{4}{|c|}{AHI error (events/hour)} \\\hline
    1 & 0.8 & 4.0 & 0.7 \\
    2 & 1.7 & 2.9 & 3.3 \\
    3 & 6.0 & 2.6 & 1.3 \\
    4 & 4.8 & 0.3 & 0.9 \\
    5 & 4.9 & 0.9 & 2.6 \\
    6 & 11.0 & 7.4 & 6.0 \\
    7 & 7.7 & 7.9 & 0.9 \\
    \hdashline
    Ave. & 5.3 & 3.7 & 2.2 \\\hline
    \multicolumn{4}{|c|}{$\mathrm{F}_1$ score} \\\hline
    1 & 0.58 & 0.73 & 0.73 \\
    2 & 0.52 & 0.76 & 0.77 \\
    3 & 0.58 & 0.62 & 0.63 \\
    4 & 0.46 & 0.58 & 0.60 \\
    5 & 0.20 & 0.23 & 0.34 \\
    6 & 0.50 & 0.59 & 0.61 \\
    7 & 0.24 & 0.26 & 0.41 \\
    \hdashline
    Ave. & 0.44 & 0.54 & 0.58 \\\hline
  \end{tabular}
  \label{tab:ahi}
\end{table}

\subsubsection{Use of patient-specific thresholds}
Next, we evaluate the performance of the proposed method when the threshold is individually adjusted for each patient by optimizing the threshold for each patient such that it maximizes the F$_1$ score of each patient.
This is henceforth referred to as proposed method B.
The optimal thresholds were found to be 0.65, 0.70, 0.70, 0.47, 0.32, 0.65, and 0.96 for patients 1,2,3,4,5,6, and 7, respectively.
The performance metrics of the proposed method with the optimal threshold are given in Table~\ref{tab:ahi}.
The average AHI error of the proposed method was 2.2 per hour, representing an improvement by a factor of 2.4 over the conventional method.
The average F$_1$ score improved to 0.58, representing an improvement by a factor of 1.3 over the conventional method.

In this study, the optimal threshold is calculated using the reference data from PSG.
As a preliminary step toward automatic selection of the optimal threshold, we discuss the optimal threshold for each patient.
For example, patient 7 has a high proportion of central sleep apnea (CSA) rather than obstructive sleep apnea (OSA) or mixed sleep apnea (MSA), and the optimal threshold for patient 7 was as high as 0.96, which may indicate that the optimal threshold appears to depend on the specific type of apnea that the patient suffers from.
To quantitatively examine the relationship between the optimal threshold and the types of apnea, we calculate coefficients of correlation between the proportions of the four types of apnea (OSA, CSA, MSA, and hypopnea) in all apnea events and the optimal threshold for each patient.
The correlation coefficients are found to be $\rho_\mathrm{OSA}=-0.29$, $\rho_\mathrm{CSA}=0.87$, $\rho_\mathrm{MSA}=-0.01$, and $\rho_\mathrm{Hyp}=-0.68$, indicating a positive correlation between the proportion of CSA and the optimal threshold.
Note that CSA differs from other types of apnea in that there is no respiratory effort, resulting in minimal displacement during apnea events.
Therefore, CSA is easier to detect, and the average values of the label $\hat{L}(t)$ tend to be high.
Consequently, using a higher threshold can reduce the number of false positives, suggesting that the use of a higher threshold for patients with CSA will enable more accurate detection.

\section{Conclusion}
This study proposed a radar-based method for detecting sleep apnea by applying an EM algorithm-based approach to overlapping data segments.
By averaging the detection results over multiple time intervals, the method generates a continuous output that represents the probability of apnea.
This approach mitigates the negative effect of irregular body movements that occur after apnea events.
The effectiveness of the proposed method was verified through experiments involving seven patients with SAS symptoms.
The proposed method improved accuracy in estimating the number of apnea events per hour, as well as in estimating the duration of each apnea event.
Moreover, the results suggest that the optimal binarization threshold depends on the type of apnea.
These findings demonstrate that the proposed method is effective in the non-contact and accurate detection of sleep apnea events.

\section*{Ethics Declarations}
This work involved human subjects or animals in its research.
Approval of all ethical and experimental procedures and protocols was granted by the Kyoto University, Graduate School of Medicine’s Ethics Committee under Application No. R2042.
\section*{Acknowledgment}
\addcontentsline{toc}{section}{Acknowledgment}
This work was supported in part by the SECOM Science and Technology Foundation; in part by Japan Science and Technology Agency under Grant JPMJMI22J2 and Grant JPMJMS2296; in part by the Japan Society for the Promotion of Science KAKENHI under Grant 21H03427, Grant 23H01420, and Grant 23K26115; and in part by the New Energy and Industrial Technology Development Organization.
We thank Glenn Pennycook, MSc, from Edanz (https://jp.edanz.com/ac) for editing a draft of this manuscript.


\begin{thebibliography}{16}
\bibitem{sleepapnea}
Benjafield AV {\em et al.} (2019),
``Estimation of the global prevalence and burden of obstructive sleep apnea: a literaturebased analysis,''
\textit{Lancet Respir. Med.}, vol. 7, no. 8, pp. 687--698.

\bibitem{sleep}
Budhiraja R, Budhiraja P, Quan SF (2010),
``Sleep-disordered breathing and cardiovascular disorders,''
\textit{Respir. Care}, vol. 55, no. 10, pp. 1322-–1332.

\bibitem{complicate}
Yaggi HK {\em et al.} (2005),
``Obstructive sleep apnea as a risk factor for stroke and death,''
 \textit{N. Engl. J. Med}, vol. 353, no. 19, pp. 2034–-2041.

\bibitem{card}
Yoon H {\em et al.} (2018),
``Sleep-dependent directional coupling of cardiorespiratory system in patients with obstructive sleep apnea,''
 \textit{IEEE Trans. Biomed. Eng.}, vol. 65, no. 12, pp.2847--2854.

\bibitem{PSG}
Kapur VK {\em et al.} (2017),
``Clinical practice guideline for diagnostic testing for adult obstructive sleep apnea: an American Academy of Sleep Medicine clinical practice guideline,''
 \textit{J. Clin. Sleep Med.}, vol. 13, no. 3, pp. 479–-504.

\bibitem{nonmachine1}
Kagawa M, Tojima H, Matsui T (2014),
``Non-contact screening system for sleep apnea-hypopnea syndrome using the time-varying baseline of radar amplitudes,''
 \textit{2014 IEEE Healthc. Innov. Conf. (HIC)}, pp.99–-102.

\bibitem{nonmachine2}
Kan S {\em et al.} (2020),
``Non-contact diagnosis of obstructive sleep apnea using impulse-radio ultra-wideband radar,''
 \textit{Sci. Rep}, vol. 10, no. 1, pp. 5261.

\bibitem{nonmachine3}
Nakajima R, Taki K, Wang H, Ma J (2024),
``Real-time respiratory apnea detection using mmwave radar,''
\textit{2024 IEEE Smart World Congr. (SWC)}.

\bibitem{machine1}
Islam SMM {\em et al.} (2020),
``Identity authentication of OSA patients using microwave Doppler radar and machine learning classifiers,''
 \textit{2020 IEEE Radio and Wireless Symp. (RWS)}, pp. 251–-254.

\bibitem{machine2}
Koda T {\em et al.} (2021),
``Radar-based automatic detection of sleep apnea using support vector machine,''
\textit{2020 Int. Symp. Antennas Propag. (ISAP)}, pp. 841–-842.

\bibitem{machine3}
Kwon HB {\em et al.} (2021),
``Hybrid CNN-LSTM network for real-time apnea-hypopnea event detection based on IR-UWB radar,''
\textit{IEEE Access}, vol. 10, pp. 17556--17564.

\bibitem{machine4}
Snigdha F, Islam SMM, Boric-Lubecke O, Lubecke V (2020),
``Obstructive sleep apnea (OSA) events classification by effective radar cross section (ERCS) method using microwave doppler radar and machine learning classifier,''
 \textit{2020 IEEE MTT-S Int. Microwave Biomed. Conf. (IMBioC)}.

\bibitem{EM}
Koda T {\em et al.} (2024),
``Noncontact detection of sleep apnea using radar and expectation-maximization algorithm,''
\textit{IEEE Sensors J.}, vol. 24, no. 20, pp. 32748--32756.

\bibitem{length}
Leppänen T {\em et al.} (2016),
``Length of individual apnea events is increased by supine position and modulated by severity of obstructive sleep apnea,''
 \textit{Sleep Disord.}, vol. 1.

 \bibitem{arousal}
Stradling JR {\em et al.} (1999),
 ``Variation in the arousal pattern after obstructive events in obstructive sleep apnea,''
 \textit{Am. J. Respir. Crit. Care Med.}, vol. 159, pp. 130--136.
\end{thebibliography}
\end{document}